\documentclass[newversion]{aa}

\usepackage{graphicx}
\usepackage{txfonts}
\usepackage{booktabs}

\usepackage{natbib}
\usepackage[utf8]{inputenc}
\usepackage{multirow}
\bibpunct{(}{)}{;}{a}{}{,} 

\newcommand{\swift}{\textit{Swift}}

\usepackage[usenames,dvipsnames]{color}
\usepackage{colortbl,xcolor}

\begin{document}

\title{GRB 090426: Discovery of a jet break in a \\
 short burst afterglow}

\author{
A.~Nicuesa Guelbenzu\inst{1}, 
S.~Klose\inst{1},
A.~Rossi\inst{1},
D.~A.~Kann\inst{1},
T.~Kr\"uhler\inst{2,3},
J.~Greiner\inst{2},
A. Rau\inst{2},
F.~Olivares~E.\inst{2},
P.~M.~J.~Afonso\inst{2}\fnmsep\thanks{\emph{Present address: }
American River College, Department of Physics and Astronomy, 4700 College Oak Drive,
Sacramento, CA 95841, USA},
R.~Filgas\inst{2},
A.~K\"upc\"u~Yolda\c{s}\inst{4},
S.~McBreen\inst{5},
M. Nardini\inst{2},
P.~Schady\inst{2},
S.~Schmidl\inst{1},
A.~C.~Updike\inst{6,7,8},
\and
A.~Yolda\c{s}\inst{4}
}

\institute{Th\"uringer Landessternwarte Tautenburg, Sternwarte 5, D--07778 Tautenburg, Germany\\
\email{ana@tls-tautenburg.de}
\and
Max-Planck-Institut f\"ur Extraterrestrische Physik, Giessenbachstra\ss{}e, D--85748 Garching, Germany
\and
Universe Cluster, Technische Universit\"{a}t M\"{u}nchen, Boltzmannstra\ss{}e 2, D--85748, Garching, Germany
\and
Institute of Astronomy, University of Cambridge, Madingley Road CB3 0HA, Cambridge, UK
\and
School of Physics, University College Dublin, Dublin 4, Republic of Ireland
\and
Clemson University, Department of Physics and Astronomy, Clemson, SC 29634-0978, USA
\and {
CRESST and the Observational Cosmology Laboratory, NASA/GSFC, Greenbelt, MD 20771, USA 
\and
Department of Astronomy, University of Maryland, College Park, MD 20742, USA}}

\date{Received 2011 February 4; accepted 2011 April 16}

\authorrunning{Nicuesa Guelbenzu et al.}
\titlerunning{GRB 090426}

\abstract
{The link between the duration of GRBs and the nature of their progenitors remains disputed.
Short bursts (with durations of less than $\sim$2\,s)
are less frequently observed, technically more difficult to
localize, and exhibit significantly fainter afterglows.}
{It is of critical importance to establish  whether the burst duration can reliably  distinguish 
the different GRB population models of collapsars and compact stellar mergers.
The \swift\ GRB~090426 provides an unique
opportunity to address this  question. Its duration ($T_{90}=1.28$\,s)
places GRB~090426 firmly in the short  burst population, while the high
redshift ($z=2.609$), host galaxy properties, and prompt emission  spectral
characteristics are more similar to those of long-duration GRBs.
}
{On the basis of data obtained with the Tautenburg 2m telescope (Germany) and  the
7-channel imager GROND (La Silla, Chile), we compiled the most finely sampled light curve
available for a
short burst optical/NIR  afterglow. The light curve
was then analysed in a standard fashion.  GROND and XRT data were used to
determine the broad-band spectral energy distribution of  the afterglow across
more than three orders of magnitude.}
{Our data show that a light curve break exists at 0.4 days,
which is followed by a steep decay. This light curve decay is achromatic in the 
optical/NIR bands, and interpreted as a post-jet break phase.
The X-ray data do not disagree with this interpretation.}
{The half-opening angle of the suspected jet as well as the luminosity
of the optical afterglow  provide additional  evidence that
 GRB~090426  is probably linked to the death of a massive star rather than to
the merger of two compact objects. }
{}

\maketitle

\section{Introduction}

{ It is} commonly accepted that
long GRBs are linked to the core-collapse of massive stars (so-called
collapsar events; \citealt{WB2006}) residing in  star-forming galaxies,  while
short bursts are linked to compact stellar mergers in all morphological types
of galaxies { (\citealt{Fong2010,Nakar2007}).}  Since the launch of the
\swift\ satellite  (\citealt{Gehrels2004}), about 100 long GRBs have been
rapidly localized per year { (\citealt{Gehrels2009}).}  Nearly half of them
have a  detected optical afterglow and one third have redshift determinations
{(see J. Greiner's web-page\footnote{http://www.mpe.mpg.de/$\sim$jcg/grbgen.html})}.  
Compared to the long
burst sample, short bursts are less frequently observed, and generally
followed by  on average significantly fainter and less luminous afterglows
(\citealt{Kann2008,Kann2010,Nysewander2009}). By the end of 2010, about three
dozen short GRBs had been localized by \swift\ (fewer than 10 events per
year). Among them, about 50\% have optical detections and about one third have
redshift determinations based on host galaxy spectroscopy
(\citealt{Berger2010}). 

GRB 090426 is an outstanding short burst ($T_{90}\sim$ 1.28 s), because it has
by far the highest redshift known among the short burst sample ($z=2.609$;
\citealt{Antonelli2009,Levesque2010}). All other short bursts with secure
redshift measurements have $z\lesssim1.1$ (\citealt{Berger2010}). The redshift
of GRB 090426 is therefore in much closer agreement with the distribution of
long GRB redshifts than with that of the  short burst sample. Several
arguments have been put forward that GRB 090426 is not related to merging
compact objects.  \citet{Antonelli2009} found that the GRB spectral and energy
properties are more comparable to those of  collapsar
events. \citet{Levesque2010} noted that the blue star-forming host of this
burst might also be consistent with a collapsar origin. Similarly, \citet{Xin2010} argued
that the deduced lower limit to the circumburst particle number density (about
10~cm$^{-3}$) is much higher than expected for a merging  stellar system,
being more characteristic of a star-forming region. 

We present additional multi-color photometry of the
optical/NIR afterglow of GRB 090426 from about 0.3 to 2.5 days 
after the burst, showing that the optical light
curve has a well-defined break at late times.\footnote{{ In the following
we use the standard notation for the flux density of the afterglow, 
$F_\nu(t) \propto t^{-\alpha}\, \nu^{-\beta}$.}} 

\section{Observations and data reduction}

Observations of the optical/NIR 
afterglow of GRB 090426 were performed with the 2m
telescope of the Th\"uringer Landessternwarte Tautenburg (TLS, Germany) and
the 7-band multichannel imager GROND (\citealt{Greiner2008}) mounted at the
2.2m ESO/MPI telescope on La Silla (Chile).

Tautenburg started observing in the $I_c$ and  $R_c$ band at 19:58 UT on 26
April 2009  and stayed on the field for 90 min. The average airmass was 1.1
and the average  seeing 1.2 arcsec.  About 3.5 \,hr after the end of the TLS
observations, at 01:08 UT on 27 April  2009 (12.3 \,hr after the GRB trigger),
GROND started following the afterglow once the target became visible over La
Silla.  Observations continued until 04:55 UT at an average seeing of 1.2
arcsec and an average airmass of 2.5, during which GROND was able to detect
the afterglow in 10 different OBs\footnote{technical name for a pre-defined
observing sequence} (Table~\ref{tab:logGROND}).  Second and third epoch
observations were obtained the following two nights. 

Data reduction was performed in a standard fashion.
TLS and GROND  data were analysed through standard PSF photometry using
DAOPHOT and ALLSTAR tasks of IRAF (\citealt{Tody1993}), in a similar way to the
procedure described in \citet{Thomas2008} and \citet{Aybuke2008}. Calibrations
were performed against the SDSS, using the transformation equations of
Lupton\footnote{ http://www.sdss.org/dr7/algorithms/sdssUBVRITransform.html}
for  the TLS data.  Magnitudes were corrected for Galactic extinction, assuming
$E(B-V)=0.017$ mag \citep{Schlegel1998} and a ratio of total-to-selective
extinction of $R_V=3.1$.

\begin{figure}[!tb]
\includegraphics[width=\columnwidth,angle=0]{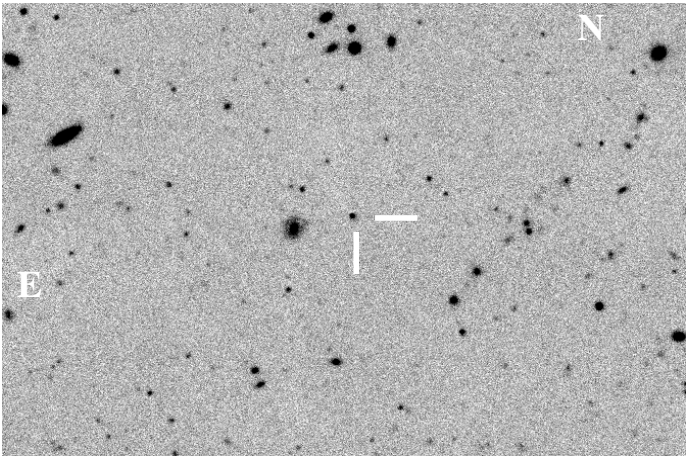}
\caption{Finding chart of the afterglow of GRB 090426. { The $r^{\prime}$-band 
GROND image is a combination of OBs 6 to 10 (Table~\ref{tab:logGROND}). The
field of view is approximately $2'\,\times\,3'$}.}
\label{fig:fchart}
\end{figure}

\section{Results}

\subsection{Spectral energy distribution (SED) \label{grondsed}}

\begin{figure}[t]
\includegraphics[width=\columnwidth,angle=0]{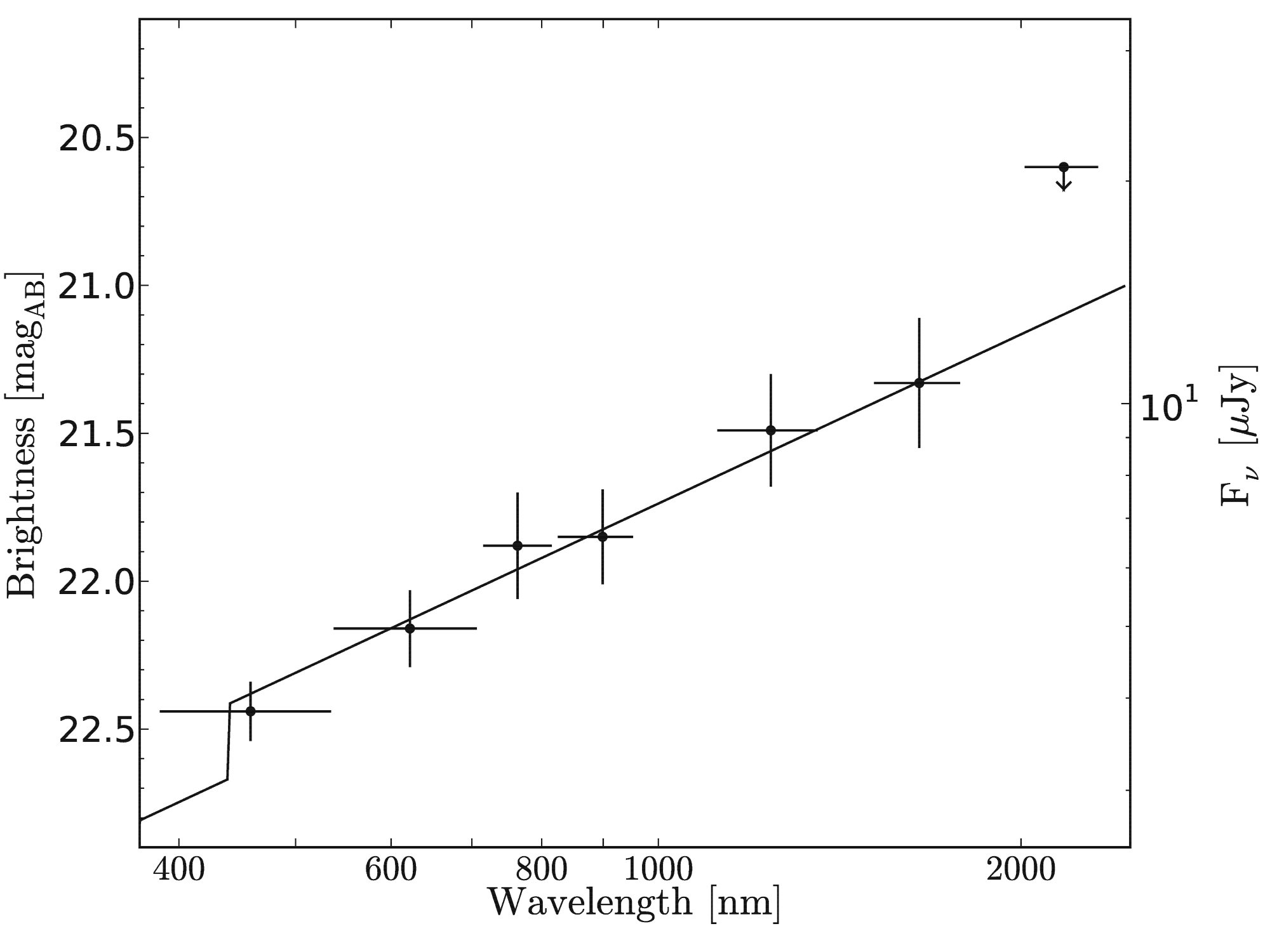}
\caption{The SED of the afterglow at a mean time of $t$=53~ks
(Table~\ref{tab:logGRONDNIR}). The data is corrected
for Galactic extinction and underlying host galaxy contribution. The fit is a
single power law ($\chi^2/$d.o.f. = 0.17). There is no evidence of extinction
by dust in the GRB host galaxy ($A_V$(host)=0 mag). 
For the fit, the redshift was fixed to $z=2.61$, { the Lyman~$\alpha$
absorption affects the $g'$ band slightly.} From left to
right, we present results for the $g^{\prime}r^{\prime}i^{\prime}z^{\prime}JHK$ bands.}
\label{fig:beta}
\end{figure}

{ The host galaxy of GRB 090426 is an extended source (about 2$''$) and the
afterglow lies above the brightest part of this galaxy, the N-E knot
\citep{Antonelli2009}. For the galaxy as well as this knot, $g'r'i'z'$
magnitudes were previously published (\citealt{Antonelli2009}), but in the
NIR bands only upper limits are known
(\citealt{Levesque2010}). Therefore,  for the construction of the 
afterglow SED, the GROND optical bands could be corrected for the
contribution of the underlying host galaxy flux, but in the NIR only the maximum
possible contribution of  host galaxy light could be considered.

Combining the GROND OBs 6 to 10 provides a good signal-to-noise ratio for the
detection of the optical transient (afterglow plus underlying host galaxy) in
$g^{\prime},r^{\prime},i^{\prime},z^{\prime},J$, and $H$
(Table~\ref{tab:logGRONDNIR}). Given the upper limits  to the $J,H$-band
magnitudes of the underlying host, the contribution of host galaxy flux in $J$
and $H$ at this time translates into an increase in brightness 
of the optical transient by at most 0.1 mag and 0.13 mag,
respectively. To correct for this contribution,
we added 0.1 mag to the $J$ as well as $H$-band data point and also
increased the corresponding 1$\sigma$ error by 0.1 mag. Performing the fit
from $g'$ to $H$ then gives $\beta_{\rm opt/NIR} = { 0.76 \pm 0.14}$
(Fig.~\ref{fig:beta}), which is close to the observed mean for optical/NIR
afterglows (about 0.6; cf. \citealt{Greiner2011,Kann2010}).}

\subsection{{ Multi-color} light curve \label{lc}}

\begin{figure}[tb]
\includegraphics[width=9.0cm,angle=0]{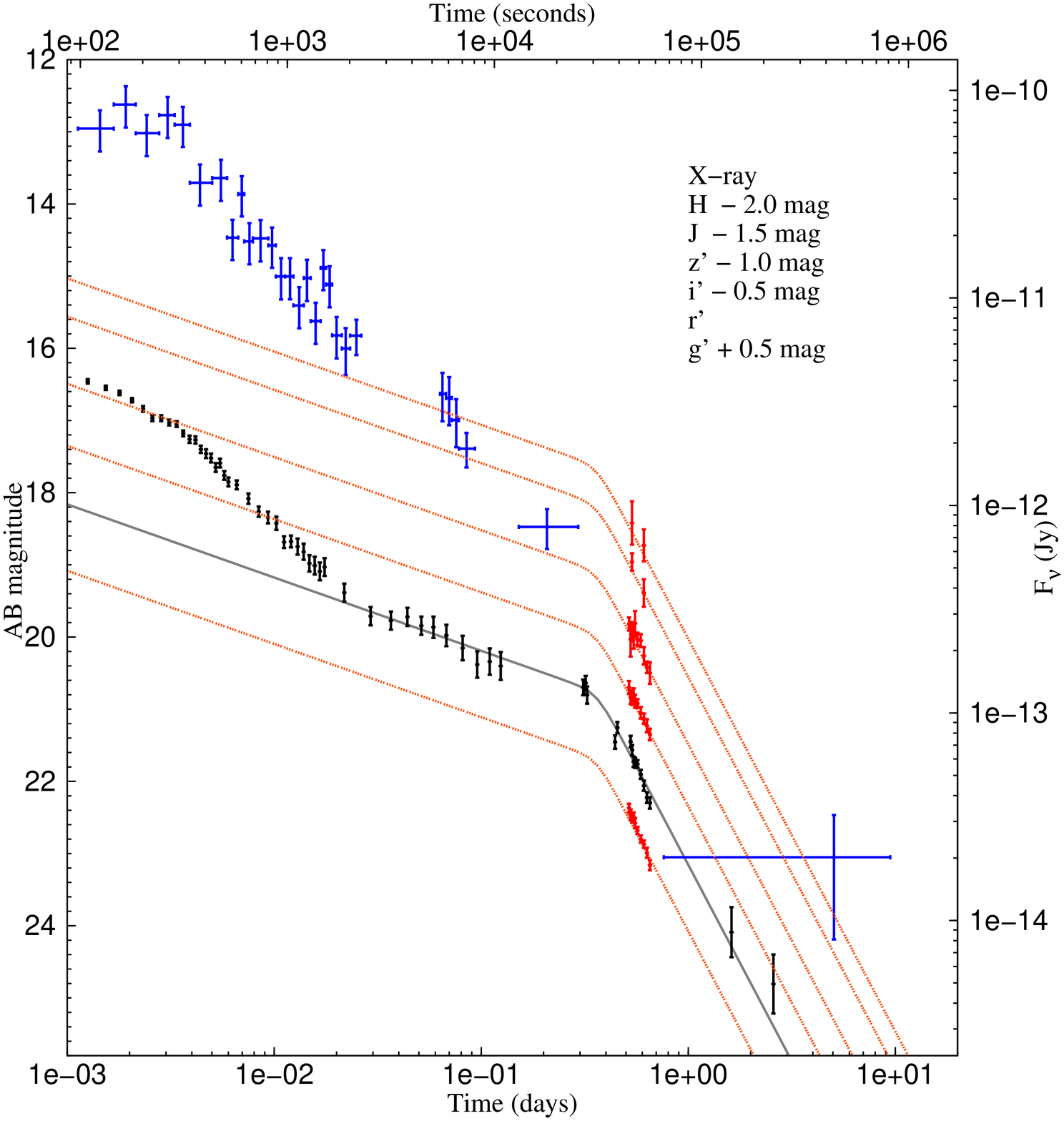}
\caption{
$R_c$-band light curve of the GRB 090426 afterglow and best  fit 
with a single broken power law after host galaxy subtraction
and correction for Galactic extinction. The fit
starts at 0.05 days.  For comparison, the $g^{\prime}i^{\prime}z^{\prime}JH$
bands (left $y$-axis; Tables~\ref{tab:logGROND},
\ref{tab:logGRONDNIR}, \ref{tab:logTLS})
and the X-ray light curve (0.3 to 10 keV, 
\citealt{Evans2010}; right $y$-axis) are also shown.}
\label{fig:lc}
\end{figure}

We combined TLS/GROND data with those of   \cite{Xin2010},
\cite{Levesque2010}, and \cite{Antonelli2009}  to obtain an $R_c$-band light
curve composed of data sets published in refereed papers.  Assuming a
power-law SED of the afterglow and a non-evolving spectral index,  the
$r^{\prime}$-band data  were transformed into $R_c$.  For completeness, the
$V$-band data from \citet{Xin2010} was also used and shifted into the $R_c$
band.  The final $R_c$-band data set after the first break at 
about 0.05 days can be fit using a single broken power law
(\citealt{Beuermann1999}; Fig.~\ref{fig:lc}). 

{ Fitting the data we find} a late break in the light curve  at around 0.4 days,
in addition
to a first break at around 0.02 days that had been previously known. 
This second break in the light curve was not seen in the
previous data sets of GRB 090426. The
$g^{\prime}r^{\prime}i^{\prime}z^{\prime}JH$ GROND data show that this
evolutionary phase is achromatic within the optical/NIR bands
(Fig.~\ref{fig:lc}). { The most obvious interpretation is that this 
is a jet break.}

\section{Discussion}

\subsection{{ The light curve parameters} \label{sectjet} }

{ Jet} breaks are  widely studied features of long GRB afterglow light
curves { (cf. \citealt{Frail2001}).} In their long burst sample,
\cite{Zeh2006} find a nearly log-normal distribution of jet  half-opening
angles between 2 and 12 degrees, with the peak around 2 to 4 degrees (see also
\citealt{Racusin2009}).  For short GRBs, however, afterglow light curves are
typically sparsely sampled. Some cases seem uncollimated \citep{Grupe2006},
while others display breaks and steep late slopes, which are evidence of
collimation { \citep{Burrows2006,McBreen2010,Soderberg2006}.}  Most of
these results, however, rely on the corresponding X-ray light curve.  { In
the case of GRB 090426, on the other hand, there are basically 
no X-ray data available around the
time of the late break in the optical light curve.} The available data
seem to indicate a smooth X-ray
afterglow decay from 4000~s on  up to the last
X-ray detection at about $4\,\times\, 10^5$~s (\citealt{Xin2010}). 
{ However, a
break in the X-ray light curve  at 0.4 days, which is simultaneous with the break in
the optical bands, is not ruled out.}

\begin{figure}[tb]
\includegraphics[width=9.1cm,angle=0]{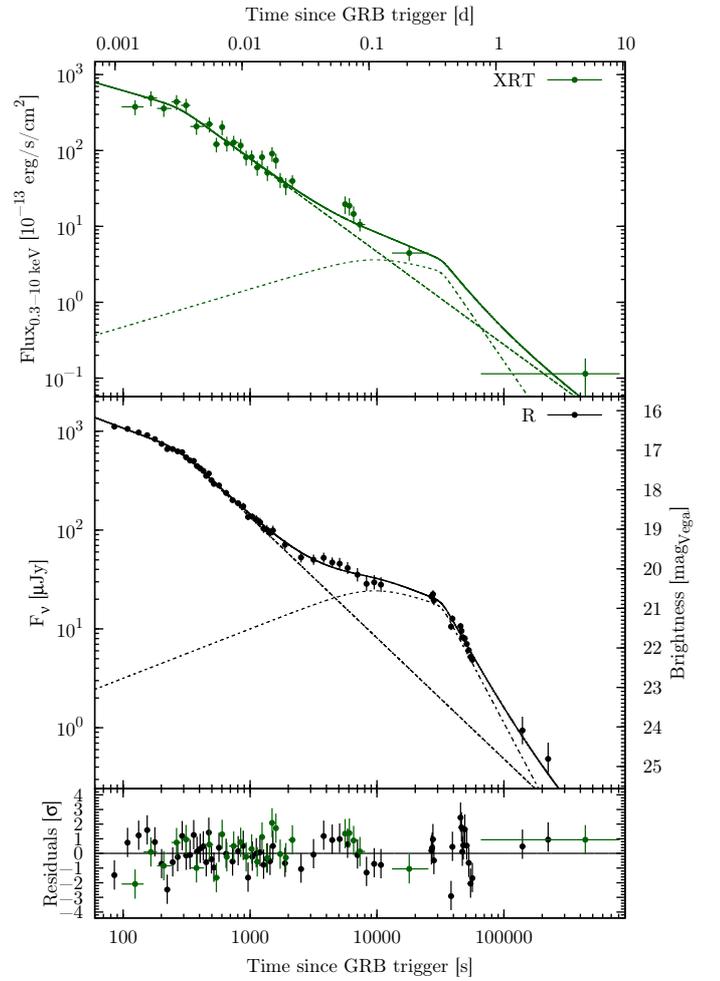}
\caption{Best fit of the optical afterglow using a two-component jet model. 
{\it Top:} X-ray light curve (0.3 to 10 keV; \citealt{Evans2010}). {\it Bottom:}
Combined $R_c$-band data set { (the same as in Fig.~\ref{fig:lc}).}}
\label{fig:2jet}
\end{figure}

{ 
A satisfying fit of the entire  optical and X-ray light curve
can be obtained by assuming a two-component jet model}
(\citealt{Berger2003,Peng2005,Racusin2008,Filgas2011};  Fig.~\ref{fig:2jet}).
Within this framework,  the observed afterglow light curve is the
superposition of the radiation from two jets, a narrow and a wide jet. { 
Even though it was not the aim of this paper to explore the
validity of this model for GRB 090426, we used it to fit the
data and to shift X-ray and optical
data points to the same time after the burst in order to
obtain the SED from the optical to the X-ray band.}

According to the best fit, the narrow-jet component is described by a
single broken power law with $\alpha_1 = 0.48\pm 0.04, \alpha_2 = 1.22\pm
0.05$, and a break time $t_{b1} = 290\pm 20$~s (while fixing the smoothness
parameter $n_1$ to 3). The second, wider component, follows a
double broken power law with $\alpha_4 = 0.46 \pm 0.15, \alpha_5 = 2.43\pm
0.19$ and  break times $t_{b2} = 9400 \pm 3800$~s ($0.11 \pm 0.04$ days) 
and $t_{b3} = 34500 \pm 1800$~s ($0.39 \pm$
0.02~days; by requiring $\alpha_3 = -0.5$ { [\citealt{Panaitescu2000}],} 
$n_2$=10, and $n_3$=10; $\chi^2$/d.o.f =  90.39/78 = 1.16).
{ The optical and the X-ray light curve trace each other, implying that 
both are belonging to the same spectral regime. We caution, however, that 
after about 0.2 days the X-ray data do not constrain the 
corresponding fit very much.}

The flat decay of the light curve after its first break at 
$t = t_{\rm b1}$ as well as the closure
relations (\citealt{Zhang2004}) show that this first break cannot be the jet
break of the narrow-jet component. { A plausible explanation of this
part of the light curve could the cessation of an energy injection episode 
(cf. \citealt{Zhang2006}; see also \citealt{Xin2010}). 
However, of primary interest here is only the late break of the optical
light curve at $t = t_{\rm b3}$. Regardless of the most suitable model,
a single-jet model (Fig.~\ref{fig:lc}) or a two-component jet model
(Fig.~\ref{fig:2jet}), the 
deduced break time as well as the post-break decay slope do not change.}

\subsection{The closure relations and the jet parameter} 

The \swift/XRT data (\citealt{Evans2010})  show that for $t>10^4$~s the
spectral slope in the X-ray band is constant with $\beta_X=0.91\pm0.30$. 
{ Combining this with the measured spectral slope in the
optical/NIR bands (Sect.~\ref{grondsed}), and  applying the closure relations,
it follows that  for $t>0.4$ days a jet model 
is in reasonable agreement with the observations
(Table~\ref{tab:alphabeta}). Thereby, a jet model with 
sideways expansion is preferred.} 
  
Following \cite{Sari1999}, the jet  half-opening angle 
$\Theta^{\rm ISM}_{\rm jet}$ for an ISM environment is
$\Theta^{\rm ISM}_{\rm jet} = 
\frac{1}{6}\left(\frac{t_b}{1+z}\right)^{3/8}\left(\frac{n_0\, 
\eta_\gamma}{\kappa\ E_{52}}\right)^{1/8}\,
$
,where, $E_{52}$ is the isotropic equivalent energy of the prompt emission in
units of $10^{52}$ erg, $n_0$ is the density of the ambient medium in
cm$^{-3}$, $\eta_\gamma$ is the efficiency of the shock in converting the
energy of the ejecta into gamma radiation, and $t_b$ is the break time in
days. { The parameter $\kappa$ is 1.0 for a single-jet scenario, while}
within the context of the two-component jet model the isotropic
equivalent energy of the wide jet is around 10\% of the corresponding
number for the narrow-jet component (\citealt{Peng2005}), { 
i.e. it is $\kappa =0.1$}.  Using $n_0=10$
cm$^{-3}$ (\citealt{Xin2010}) and $\eta_\gamma=0.2$, for the observed break
time at $t_b$= 0.4 days with $E_{52}=0.42^{+0.59}_{-0.04}$
(\citealt{Levesque2010}), it follows that 
$\Theta_{\rm jet}^{w} = (6.5 \pm 0.4$) degrees
and  $E_\gamma^{\rm corr, w}$ [1 to $10^4$ keV]  = $(4.2 \pm 1.4)\,\times\,
10^{48}$ erg. { Within the framework
of the single-jet scenario, the break time is basically the same 
(Fig.~\ref{fig:lc}), $\kappa=1$, and it follows that
$\Theta_{\rm jet} = (4.8 \pm 0.3$) degrees,
as well as  $E_\gamma^{\rm corr}$ = $(2.3\pm0.8)\,\times\,10^{49}$ erg.}

Before the suspected jet break at 0.4 days, the observed light curve
decay between about 0.1 and 0.4 days is quite shallow ($\alpha_4$; 
Fig.~\ref{fig:2jet}),
while the closure relations in this case, for a model with isotropic
expansion, predict a steeper decay ($\alpha>1$). We caution, however, that the
lack of data in this evolutionary phase makes it impossible to draw definite
conclusions here.

\subsection{The afterglow compared to other short bursts}

Some authors have proposed (e.g.,
\citealt{Zhang2009}) a phenomenological classification based on the link with
the progenitor, defining Type II in the case of a collapsar event and Type I in
the case of  merging compact objects, independent of the  actual duration of the
GRBs.

Using the methods detailed in \cite{Kann2006}, we created a composite light
curve of the afterglow  of GRB 090426 and compared it to the afterglow samples
of \citet[][ Type II GRBs]{Kann2010},  and \citet[][ Type I GRBs]{Kann2008}.
Observationally, the afterglow is seen to lie in the faint end of the
distribution of Type II GRB  afterglows (Fig.~\ref{Kann1}), especially in cases of
late steep decay, and there are several Type I  GRB afterglows that are
brighter  (e.g., GRB 050724, \citealt{Berger2005}; GRB
051221A, \citealt{Soderberg2006}; GRB
060614, \citealt{DellaValle2006,Fynbo2006,GalYam2006,Mangano2007}) 
at this time. However, GRB
090426 lies at a much higher redshift ($z=2.6$) than the aforementioned GRBs
($z=0.1-0.5$). In the  $z=1$ system (Fig.~\ref{Kann2}), the afterglow of GRB
090426 is clearly seen as an average  afterglow relative to the Type II
GRB afterglow sample. { At $t= 1$ day} (in the $z=1$ frame), it  would have had an
absolute  brightness of $M_B=-22.02\pm0.35$, which is just one magnitude
fainter ($2\sigma$) than the mean of the afterglow comparison sample based on
\swift \ detections \citep{Kann2010}. On the other hand, the afterglow of GRB
090426 is at  all times more luminous than any afterglow of the Type I sample
except for the controversial case  of GRB 060121 \citep{Kann2008}. There is a
strong indication that, in spite of its very short duration, GRB 090426 is a
Type II GRB, in accordance with other studies
\citep{Zhang2009,Levesque2010,Xin2010}.

\section{Summary and conclusions}

We have presented TLS/GROND data of the optical/NIR afterglow of GRB
090426, which show that the afterglow features a second break that was
missed in all previously published data sets. 
On the basis of its achromaticity in the optical/NIR bands and in agreement with the closure
relations, we have 
argued that the late light curve break at 0.4 days is a jet
break. Its calculated half-opening angle agrees well with the distribution of
half-opening angles found for long  bursts.  
In addition, the observed luminosity of
the afterglow also suggests that GRB 090426 was related to a
collapsar event.  The interesting question then is whether the short duration of
GRB 090426  in its host frame at $z=2.609$ ($T_{90}=0.33$~s) can be explained
within the framework of the collapsar model and how this compares to other
long bursts of similar  short duration in their host frame
(\citealt{Greiner2011a}). More observational data of other short burst
afterglows are needed, not only to derive more reliable statistics but also to
understand wether this short burst is an exception  rather than the rule.

\vspace{1mm}  

\noindent {\it In a recent paper, \cite{Thoene2011}
find further arguments that GRB 090426 was due to a collapsar event.}

\begin{acknowledgements}
A.N.G. acknowledges useful discussions with Francisco Molleda Sanchez (Madrid,
Spain) and Manuel Segura Morales (La Laguna, Spain). A.N.G., D.A.K., A. Rossi, 
\& S.K. acknowledges support by grant DFG Kl 766/16-1. A.N.G., A. Rossi \&
A.U. are grateful for travel funding support through MPE. A. Rossi
acknowledges support from the BLANCEFLOR Boncompagni-Ludovisi, n\'ee Bildt
foundation, T.K. by the DFG cluster of excellence 'Origin and Structure of the
Universe', F.O. funding of his Ph.D. through the DAAD, M.N. support by
DFG grant SA 2001/2-1 and P.S. by DFG grant SA 2001/1-1. Part of the funding
for GROND (both hardware and personnel) was generously granted from the
Leibniz-Prize to G. Hasinger (DFG grant HA 1850/28-1). This work made use of
data supplied by the UK Swift science data center at the University of
Leicester. We thank the referee for very helpful remarks.
\end{acknowledgements}


\bibliographystyle{aa}


\Online

\begin{appendix}

\section{Observational Data and Afterglow Luminosity} 

\begin{table}[!htb]
\caption[]{
Log of the GROND observations (in case of 
the first epoch data these are OBs 1 to 10),  with
the magnitudes given in the AB system
{ (not corrected for Galactic extinction}). 
These results supercede the data given
in \cite{Olivares2009}.}
\renewcommand{\tabcolsep}{8.5pt}
\begin{center}
\begin{tabular}{llllllll}
\toprule

 Time (s) &$g^{\prime}$        &  $r^{\prime}$      & $i^{\prime}$       &   $z^{\prime}$       &     $J$     &   $H$    &        $K$         \\
\midrule
 44729    & 21.90 (05) & 21.50 (08) & 21.46 (08) & 21.30 (08)   &  $>$21.2    &   $>$ 20.6   &   $>$19.9     \\ 
 45506    & 21.98 (06) & 21.57 (06) & 21.59 (08) & 21.50 (21)   &  $>$20.9    &   $>$ 20.2   &   $>$19.8     \\
 46268    & 22.03 (04) & 21.67 (05) & 21.59 (09) & 21.41 (11)   &  $>$21.6    &   $>$ 20.8   &   $>$19.8     \\
 47037    & 22.00 (05) & 21.82 (05) & 21.55 (08) & 21.47 (14)   &  $>$21.6    &   $>$ 21.0   &   $>$19.9     \\
 47812    & 22.08 (05) & 21.83 (04) & 21.64 (08) & 21.29 (15)   &  $>$21.7    &   $>$ 21.1   &   $>$20.2     \\
 49112    & 22.17 (03) & 21.84 (03) & 21.67 (05) & 21.50 (07)   &  $>$22.1    &   $>$ 21.4   &   $>$20.4     \\
 50930    & 22.28 (03) & 21.96 (03) & 21.79 (06) & 21.52 (07)   &  $>$22.1    &   $>$ 21.3   &   $>$20.4     \\
 52756    & 22.34 (03) & 22.10 (04) & 21.87 (05) & 21.71 (09)   &  $>$22.2    &   $>$ 21.5   &   $>$20.5     \\
 54571    & 22.44 (04) & 22.20 (03) & 21.96 (07) & 21.86 (05)   &  $>$22.0    &   $>$ 21.2   &   $>$20.5     \\
 56374    & 22.58 (04) & 22.29 (04) & 22.07 (06) & 21.93 (11)   &  $>$22.2    &   $>$ 21.6   &   $>$20.4     \\
 \midrule
 139787   & 24.03 (10) & 23.80 (11) & 23.63 (17) & 23.38 (18)   &  $>$22.7    &   $>$21.9   &    $>$21.1     \\
 222822   & 24.23 (11) & 24.24 (12) & 23.84 (15) & $>$24.1     &  $>$22.7    &   $>$21.8   &    $>$21.3     \\
\bottomrule
\end{tabular}
\label{tab:logGROND}
\end{center}
\end{table}

\begin{table}[!htb]
\caption[]{
Log of the GROND observations for the combined OBs 1 to 5 and OBs 6 to 10, with
the magnitudes given in the AB system.
{ Data are not corrected for Galactic extinction}. }
\renewcommand{\tabcolsep}{8.5pt}
\begin{center}
\begin{tabular}{llllllll}
\toprule
Time (s) &$g^{\prime}$          &  $r^{\prime}$        &    $i^{\prime}$     &   $z^{\prime}$        &     $J$      &   $H$       &        $K$     \\
\midrule
 46268 &  22.05 (03)    &  21.74 (03)  & 21.65 (05)  &  21.44 (07)   & 20.96 (02)   &  20.92 (20) &      $>$ 20.1  \\
 52723 &  22.28 (02)    &  21.99 (02)  & 21.77 (03)  &  21.62 (04)   & 21.39 (09)   &  21.23 (12) &    $>$ 20.6    \\    
    
\bottomrule
\end{tabular}
\label{tab:logGRONDNIR}
\end{center}
\end{table}

\begin{table}[!htb]
\caption[]{Log of the TLS observations, given in the Vega system.
{ Data are not corrected for Galactic extinction}.}
\renewcommand{\tabcolsep}{8.5pt}
\begin{center}
\begin{tabular}{lll}
\toprule
  Time(s)               &  $R_c$    & $I_c$        \\
\midrule
 26868    &          20.90 (13) &  --        \\
 27146    &          20.91 (14) &  --        \\
 27607    &          21.05 (19) &  --        \\
 28063    &          20.97 (14) &  --        \\
 26263   &            --        & 20.48 (18) \\
\bottomrule
\end{tabular}
\label{tab:logTLS}
\end{center}
\end{table}

\newpage
\clearpage

\begin{figure*}[htb]
\includegraphics[width=18cm,angle=0]{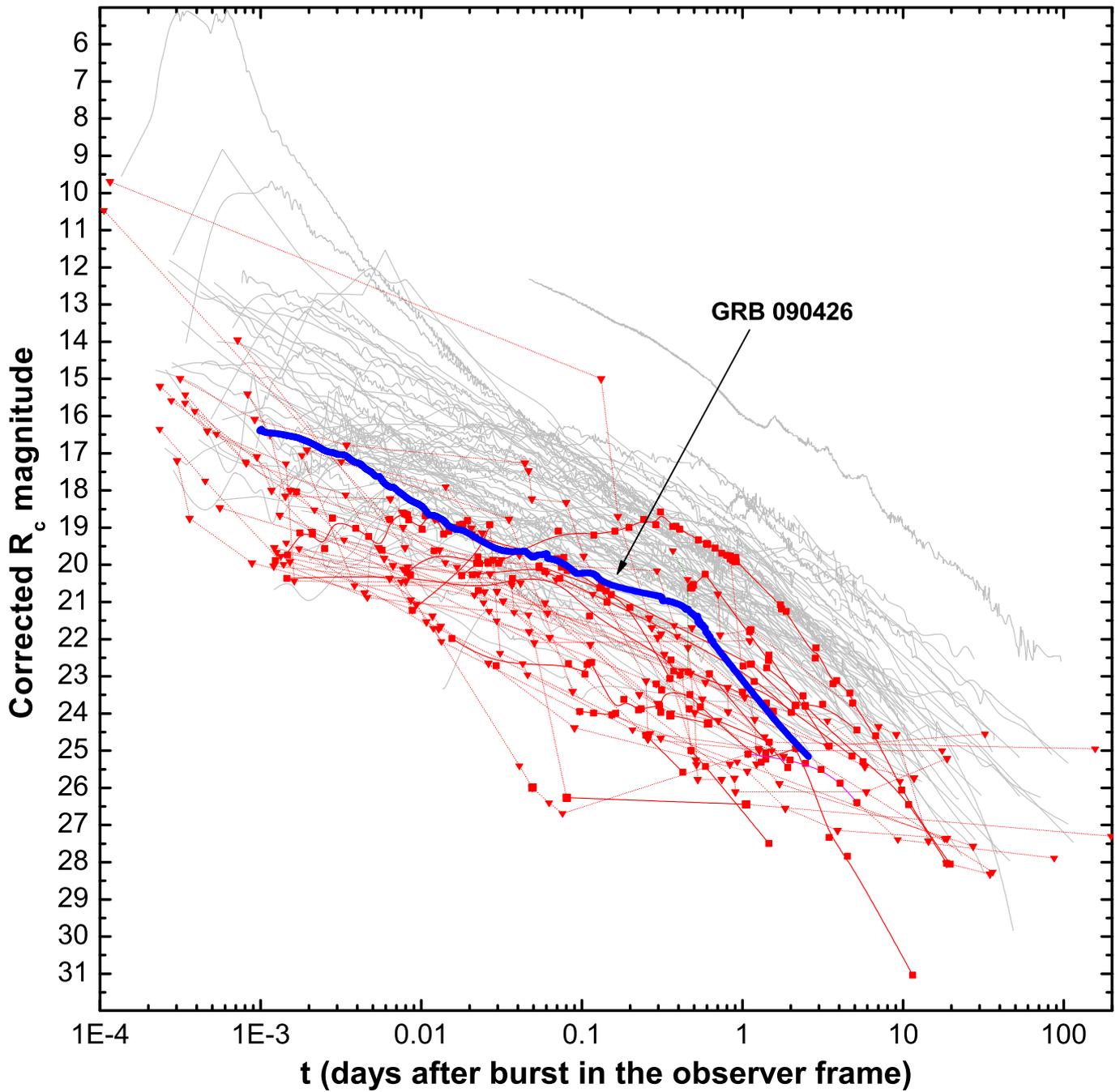}
\caption{The afterglow of GRB 090426 (thick blue line) in comparison with the
afterglows of Type II (thing gray lines) and Type I (red symbols and lines; 
squares connected by splines are detections, downward triangles connected by 
thin dashed lines are upper limits) GRBs from the sample of \cite{Kann2008,
Kann2010}. These afterglows have been corrected for Galactic extinction, and
the host galaxy contribution has been subtracted where possible (also in the
case of GRB 090426). The afterglow of GRB 090426 is seen to be among the faint
Type II GRB afterglows, but it is brighter than most Type I GRB afterglows or 
limits thereon.}
\label{Kann1}
\end{figure*}

\begin{figure*}[htb]
\includegraphics[width=18cm,angle=0]{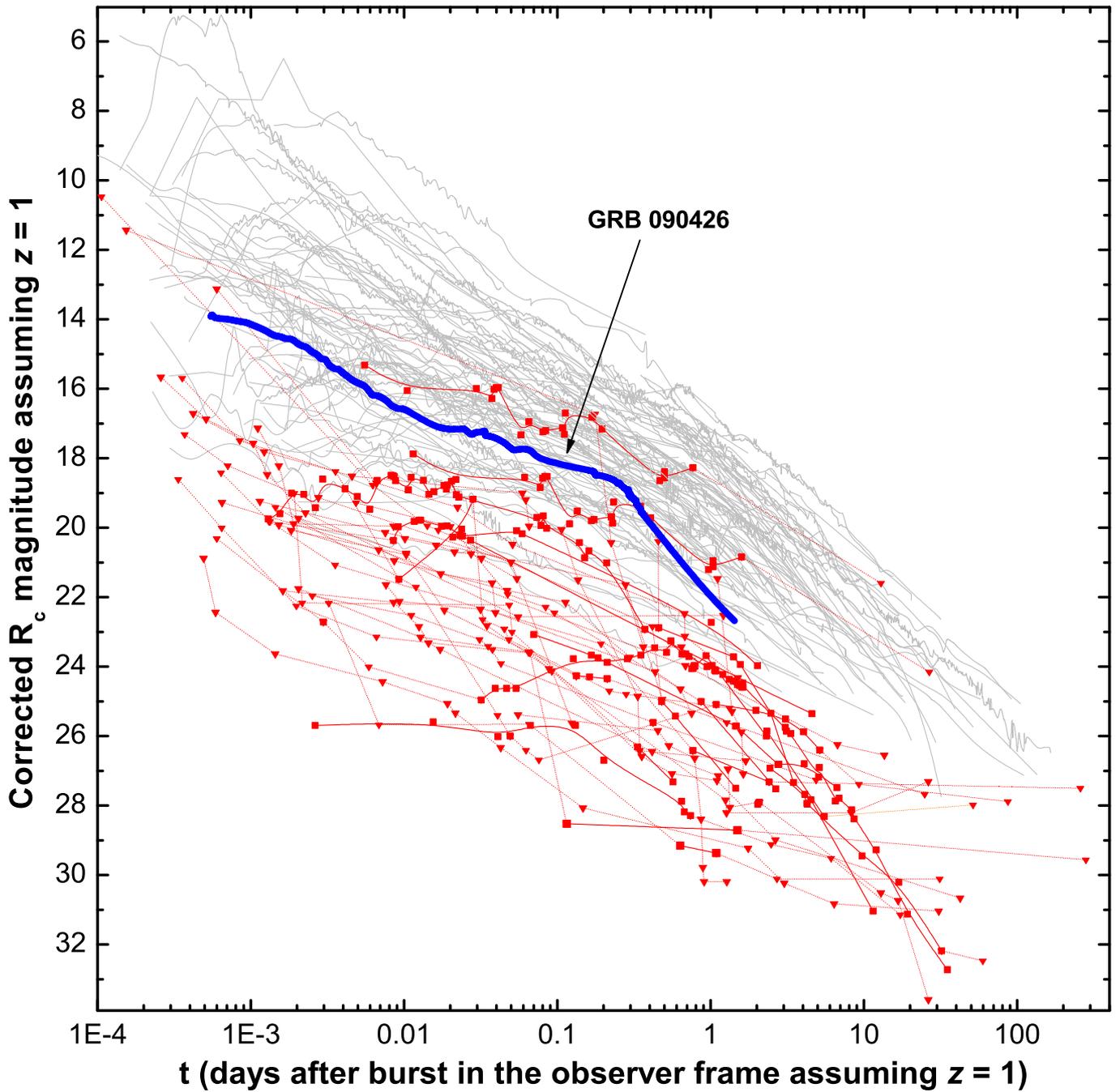}
\caption{The afterglow of GRB 090426 after it has been shifted to the $z=1$
system, again in comparison with the samples (shifted using the same method)
of  \cite{Kann2008, Kann2010}. The labelling is identical to that in
Fig.~\ref{Kann1}.  It can now clearly be seen that the afterglow of GRB
090426 is readily comparable to the afterglow of Type II GRBs (collapsar
events),  while it is much brighter than any Type I GRB  afterglow (merging
compact objects),  with the exception of GRB 060121, which is a controversial
case.}
\label{Kann2}
\end{figure*}

\begin{table}
\begin{minipage}{\columnwidth}
\renewcommand{\footnoterule}{}
\caption{
Predicted temporal decay slopes $\alpha$ for $t>0.4$ days for various
afterglow scenarios  based on the measured spectral
slopes $\beta_{\rm opt}$={ 0.76 $\pm$ 0.14} and $\beta_{\rm X} = 0.91 \pm
0.30$.}
\centering
\centering
\begin{tabular}{lccrr}
\toprule                                                                            
Model   &      Optical                 &       X-rays                         
&$s/\sigma_{\rm s,opt}$      &       $s/\sigma_{\rm s,x}$      \\
\midrule                                                                            
\multicolumn{3}{c}{Isotropic case}                       \\
ISM, $\nu<\nu_c$         &$  { 1.13}	\pm{	0.21}	$&$	1.37	\pm	0.45	$&${	-4.61}	$&$	-2.18	$\\
ISM, $\nu>\nu_c$         &$  { 0.63}	\pm{	0.21}	$&$	0.87	\pm	0.45	$&${	-6.37} 	$&$	-3.20	$\\
Wind, $\nu<\nu_c$        &$  { 1.63}	\pm{	0.21}	$&$	1.87	\pm	0.45	$&${	-2.84}	$&$	-1.16	$\\
Wind, $\nu>\nu_c$        &$  { 0.63}	\pm{	0.21}	$&$	0.87	\pm	0.45	$&${	-6.37}	$&$	-3.20	$\\
\midrule                          
\multicolumn{3}{c}{Jet with sideways expansion}             \\
\midrule                                                                            
$\nu<\nu_c$              &$  { 2.50}	\pm{	0.28}	$&$	2.82	\pm     0.60	$&${	 0.21}	$&$	0.62	$\\  
$\nu>\nu_c$              &$  { 1.50}	\pm{	0.28}	$&$	1.82	\pm	0.60	$&${	-2.75}	$&$	-0.97	$\\
\multicolumn{3}{c}{Jet without sideways expansion}           \\
\midrule                                                                            
$\nu<\nu_c$              &$  { 1.88}	\pm { 0.21}	$&$	2.12	\pm { 0.45}	$&${	-1.96}	$&${	-0.64}	$\\
$\nu>\nu_c$              &$  { 1.38}	\pm { 0.21}	$&$	1.62	\pm { 0.45}	$&${	-3.73}	$&${	-1.67}	$\\
\bottomrule       
\end{tabular}
\label{tab:alphabeta}
\footnotetext{Columns 4 and 5 give the difference between the predicted and the
observed ($\alpha= 2.43 \pm 0.19$; { the parameter $\alpha_5$ in
Sect.~\ref{sectjet})} temporal decay slope, normalized to the square root of
the sum of their quadratic errors, with $s= (\alpha_{\rm predicted}-
\alpha_{\rm observed}), \sigma_{s}^2= \sigma_{\rm predicted}^2 + \sigma_{\rm
observed}^2$.}
\end{minipage}
\end{table}

\begin{figure*}[htb]
\includegraphics[width=18cm,angle=0]{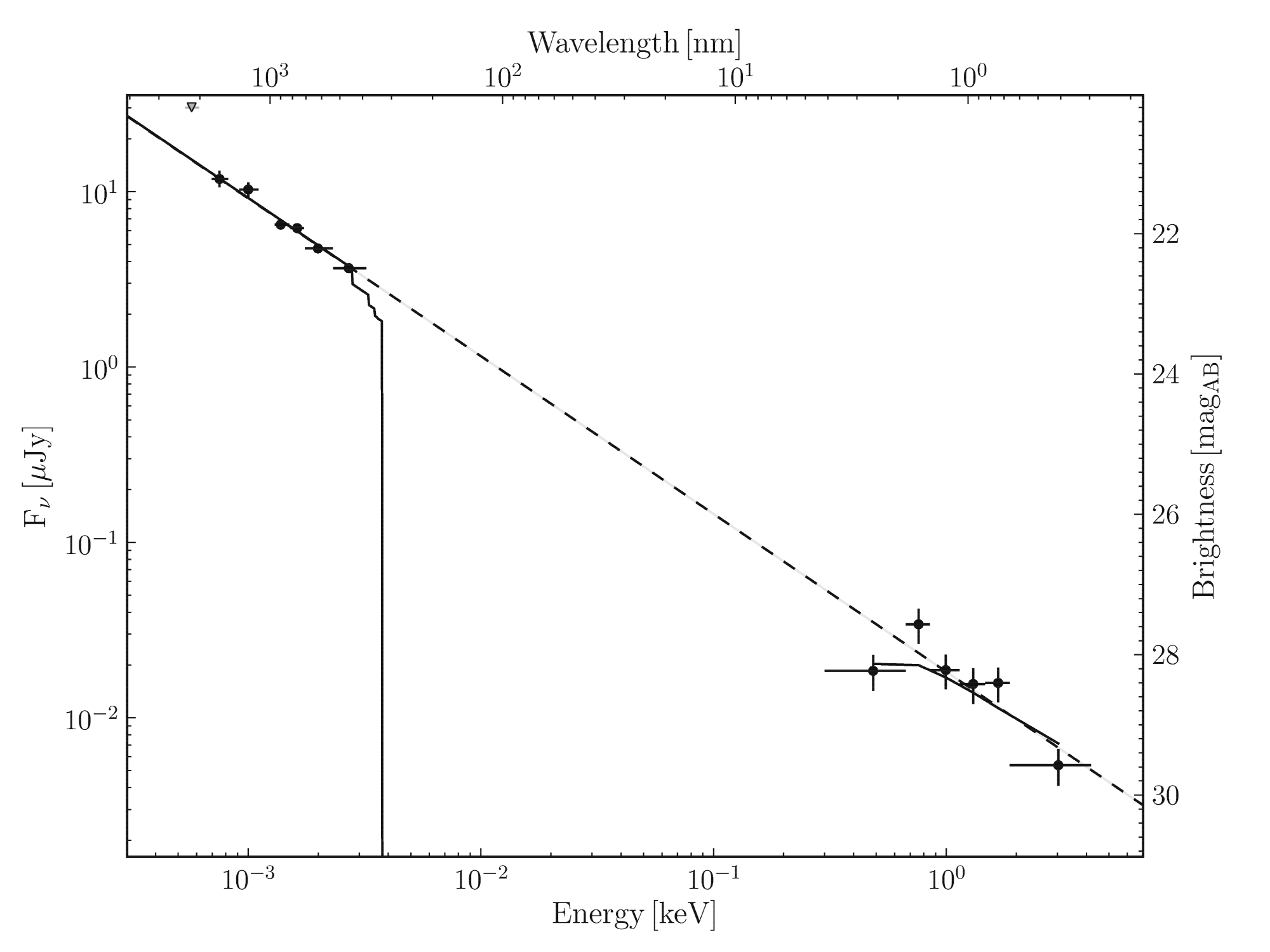}
\caption{X-ray to optical spectral energy distribution of the afterglow of
GRB 090426 at $t= 8$ks after the burst, calculated based on the 
fitted light curve (Fig.~\ref{fig:2jet}). 
The fit uses 
$N_{\rm H}^{\rm Gal} = 0.015 \,\times\, 10^{22}$ cm$^{-2}$ 
and corresponds to  a negligible  host extinction, a gas column density  of $N_{\rm
H}^{\rm host} = 0.46_{-0.46}^{0.77} \,\times\,10^{22}$ cm$^{-2}$, and a
spectral slope of $\beta_{\rm OX}=0.90 \pm 0.03$ ($\chi^2$ = 10.85 with 7
d.o.f., $\chi_\nu$ = 1.55).}
\label{fig:XOSED}
\end{figure*}

\end{appendix}

\end{document}